\documentclass[11pt,a4paper]{article}

\usepackage[T1]{fontenc}
\usepackage[utf8x]{inputenc}
\usepackage[french,english]{babel}
\usepackage{authblk}
\usepackage{graphicx}
\usepackage{algorithmic}
\usepackage{geometry}
\usepackage{amsmath,amsfonts,amssymb}
\usepackage{float}
\newcommand{\HRule}{\rule{\linewidth}{0.5mm}}
\title{Homemade assembly and parameterization of a High Performance Cluster using PelicanHPC with Flops testing and controlled temperature thanks to MCUs Arduino project}
\author{T. Maquart$^1$, G. Maquart$^2$\\$^1$Laboratory of Mechanical of Contacts and Structures (LaMCoS), CNRS, UMR 5259, INSA-Lyon,  University of Lyon, (tristan.maquart@insa-lyon.fr)\\$^2$Nuclear Physics Institute of Lyon (IPNL), Claude Bernard University Lyon 1 (UCBL), (guillaume.maquart@in2p3.fr)}
\begin{document}
\maketitle
\begin{abstract}
\noindent
This article shows a lower cost realization of a compute cluster using Debian distribution such as PelicanHPC.
We will explain parameterization and network configuration for master and compute slave nodes. Performance testing will take place using $flops.f$ file given by MPI. The results will be compared between differents clusters. We will explain quickly how the temperature is controlled by a microcontroller unit. 
\end{abstract}
\section{Introduction}
\noindent
Parallel programming is a very new tool using multiple computing cores for a given program, increasing the speedup in the parallelized instructions of the program. In this article we are particularly interested to test the performance of a cluster using the $flops.f$ file given by MPI. The cluster is based on the Debian distribution named PelicanHPC \cite{pelicanHPC} (For High Performance Clusters) with Xfce desktop environment which is lightweight (version used is PelicanHPC 3.1 \cite{down1}). Pelican is a live CD distribution that provides the ability to boot a full Linux operating system and access many software from the media (CD, DVD or USB stick) without installation on the hard disk, and without altering its content. Note that this system has great potential thanks to its support for parallel computing based on MPI using Fortran (77, 90, 95), C, C ++, but also interpreted languages ​​such as GNU Octave or semi-interpreted as Phyton.
\section{Setup}
\begin{figure}[!htb]
\centering
\includegraphics[height=.30\textheight]{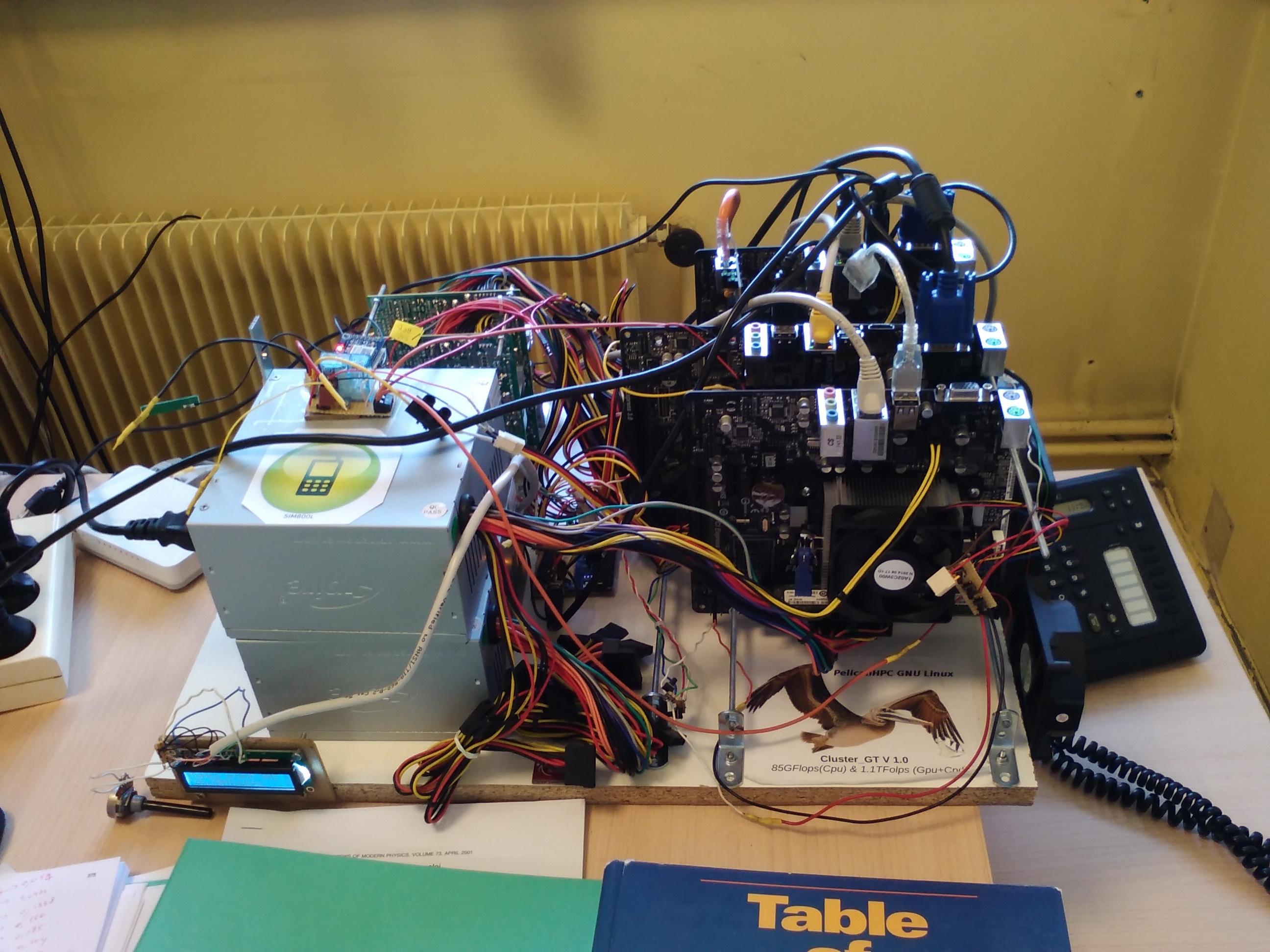}
\caption{Picture of the cluster}
\label{fig:Figure1}
\end{figure}

\begin{figure}[!htb]
\centering
\includegraphics[height=.25\textheight]{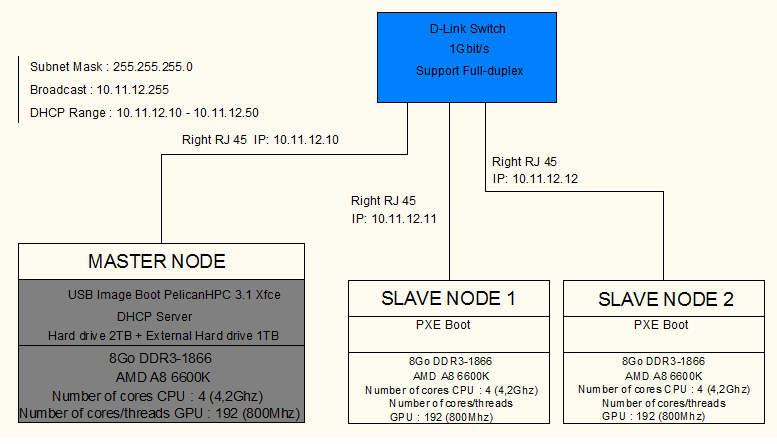}
\caption{Network configuration}
\label{fig:Figure2}
\end{figure}

\noindent
The PXE boot (acronym for Pre-boot eXecution Environment) allows a workstation to boot from the network by recovering an operating system image that is on a server \cite{tuto1}. This system image could be a gross operating system or an operating system with custom personal software components. This cluster in picture \ref{fig:Figure1} has a master node that manages the startup by PXE boot other the network of two slave nodes. As was said earlier, the PelicanHPC system is only live version, that is to say all data and configuration made on the live partition is deleted when the cluster is shutdown or restarted. Note that all nodes must be in the same subnet (see picture \ref{fig:Figure2}) to allow starting and they must have PXE option enabled in their BIOS. It may be noted that it is strongly recommended to leave PelicanHPC manage his own DHCP server to avoid conflicts. This calculation system have a hybrid memory since each of the cores of the APU AMD A8 6600K share the same memory (RAM), but the memories are different from one processor to another. Now look at the expected maximum theoretical performance:

\begin{equation}\label{eq:1}
CPU : (4,2*4*4)*3=201,6 \:\: Gflops
\end{equation}

\noindent
This theoretical performance is upper than the reality. In this calculation we have taken 4 $flops$ per cycle. It is really important to control the cooling of each processors to avoid BIOS maximum temperature during important calculations. Each processors are well-equipped by a LM35DZ temperature sensor in their heatsinks, thereafter a resolution of an unstationary differential equation of thermal conduction give the real temperature of cores. These sensors are connected to a microcontroller chip ATMEGA328P-PU with Arduino UNO bootloader for an easily acces to the open source project. An Arduino sketch is uploaded on this chip to control on/off of additional fans. A SIM800L module is connected to the MCU for sending processors temperatures by Short Message Service to directors numbers. A serial communication beetween this additional module and ATMEGA328P-PU using GSM AT commands allows the microcontroller chip to control SMS data. This chip run 350 lines of code in loop all twenty seconds. An LCD light screen shows informations about sensors. We can have more information of Arduino project by surfing on their website \cite{arduino}.

\section{Results and performance}
\begin{figure}[!htb]
\centering
\includegraphics[height=.25\textheight]{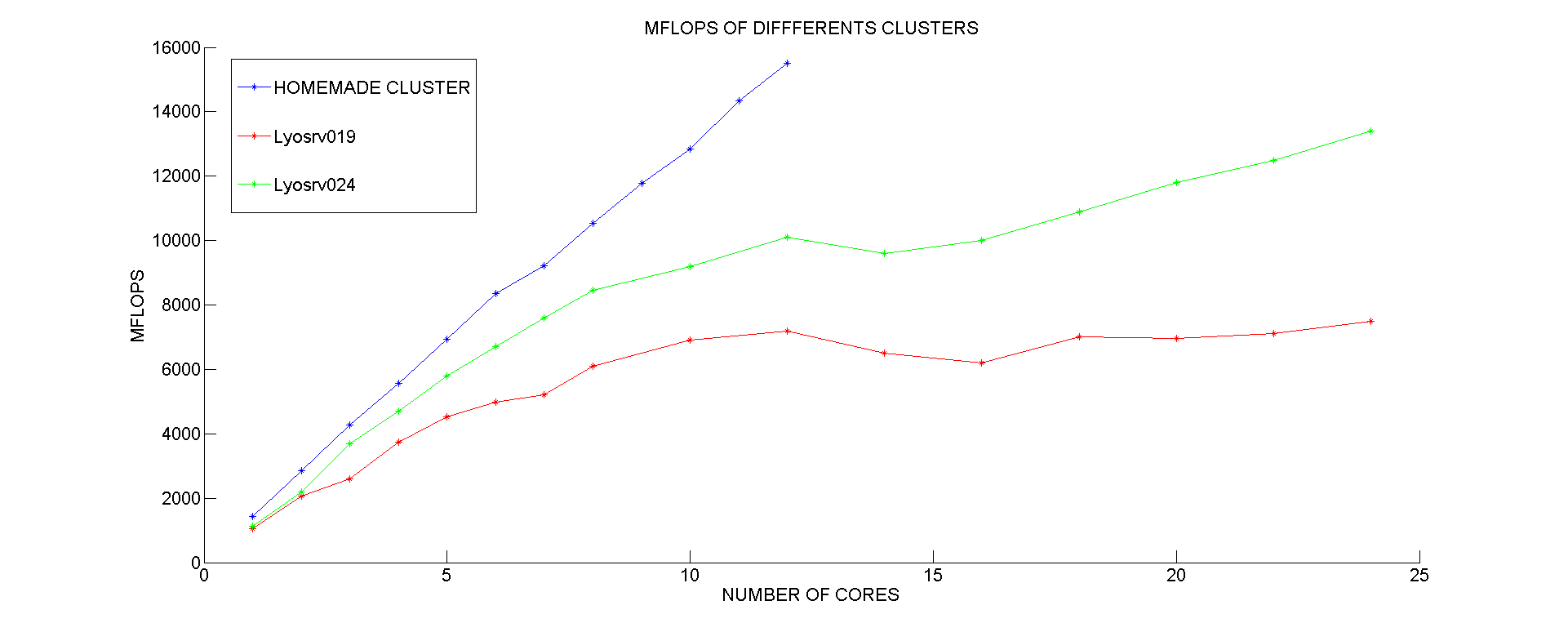}
\caption{Mflops of differents cluster versus Homemade PelicanHPC}
\label{fig:Figure3}
\end{figure}

\noindent
We will analyse quickly the results we can show on picture \ref{fig:Figure3}. There is three different cluster including our homemade PelicanHPC high performance cluster. Two others are cluster from Nuclear Physics Institute of Lyon (IPNL) respectively with 16 and 24 cores. In our small network architecture with three motherboards, we can notice that performance is almost a linear function of the number of cores. Moreover, thanks to this results, homemade 12 cores assembly appears to be more efficient than other clusters of IPNL. We will highlight that this test is running on $flops.f$ file given by MPI and it's not the real performance of your personnal parallelized code because each MPI codes uses differents physically part of clusters. However, this gives an overall idea of performance and quality of communication between computing cores. Table \ref{table:table1} lists physically description of clusters.

\begin{table}
\begin{algorithmic}
\STATE \HRule \\
\STATE \centering \textbf{Clusters quick description}
\STATE \HRule \\
\end{algorithmic}
\begin{tabular}{c|c|c}    
\begin{minipage}{6cm}
\begin{algorithmic}
\STATE Homemade Cluster
\end{algorithmic}
\end{minipage}
 &
\begin{minipage}{6cm}
\begin{algorithmic}
\STATE Lyosrv019
\end{algorithmic}
\end{minipage}
 &
\begin{minipage}{6cm}
\begin{algorithmic}
\STATE Lyosrv024
\end{algorithmic}
\end{minipage}
\end{tabular}
\begin{algorithmic}
\STATE \HRule \\
\end{algorithmic}
\begin{tabular}{c|c|c}    
\begin{minipage}{6cm}
\begin{algorithmic}
\STATE 3*AMD A8 6600K
\STATE 4 cores/CPU
\STATE 4.2 Ghz/core
\STATE 8Go DDR3-1866/CPU
\end{algorithmic}
\end{minipage}
 &
\begin{minipage}{6cm}
\begin{algorithmic}
\STATE 4*Intel(R) Xeon(R) CPU E5645  
\STATE 6 cores/CPU
\STATE 2.40 Ghz/core
\STATE 4Go (Low Frequency)/CPU
\end{algorithmic}
\end{minipage}
 &
\begin{minipage}{6cm}
\begin{algorithmic}
\STATE 4*Intel(R) Xeon(R) CPU E5-2640  
\STATE 6 cores/CPU
\STATE 2.50 Ghz/core
\STATE 16,5Go (Low Frequency)/CPU
\end{algorithmic}
\end{minipage}
\end{tabular}
\begin{algorithmic}
\STATE \HRule \\
\end{algorithmic}
\vspace{-0.3cm}
\caption{Clusters quick description}
\label{table:table1}
\end{table}
\section{Conclusion}
\noindent
It is important to know the performance of a cluster in order to adapt a parallel code. Based on certain formulations using differents parts of the cluster, the resolution will be slowest in some cases, that's why it's important to adjust code to the considered cluster configurations. To talk about price of this installation, we can have 4*3 cores clocked at 4,2 Ghz for less than 650 euros (including motherboards and network installations). The price and simplicity of implementation of this Debian distribution in a compute cluster is a real benefit not to ignore. A single cluster like this may prove more efficient than industrial cluster nodes with 16 cores or more.


\end{document}